\documentclass[12pt]{article}
\usepackage[cp1251]{inputenc}
\usepackage{amsmath,amssymb}
\usepackage[dvips]{epsfig}
\usepackage[dvips]{graphics}
\newcommand{\rhm}{\rho_{m}}
\newcommand{\prm}{p_{m}}
\newcommand{\dph}{\dot{\phi}}
\newcommand{\pat}{\partial}

\newcommand{\fl}{\relax}
\begin{document}
\begin{center}
\Large\textbf{Gravitation theory in Riemann-Cartan space-time and regular cosmology}\\
\bigskip
\normalsize A.V. Minkevich\\
\medskip
\textit{Department of Theoretical Physics, Belarussian State University, F. Skoriny av. 4, Minsk
220030, Belarus\\
Department of Physics and Computer Methods, Warmia and Mazury University in Olsztyn, Poland }
\end{center}
\begin{center}
\begin{minipage}{0.8\textwidth}
\textbf{Abstract.}  Principal ideas of gauge approach applying to gravitational interaction and
leading to gravitation theory in Riemann-Cartan space-time are discussed. The principal relations
of isotropic cosmology built in the framework of the Poincare gauge theory of gravity and the most
important their consequences are presented.
\end{minipage}
\end{center}

\section{Introduction}

As it is well known, methods of Non-Euclidian Geometry are widely applied in modern theory of
gravitation. According to Einsteinian General Relativity theory (GR) physical space-time has the
structure of 4-dimensional pseudo-riemannian continuum. GR is the base of modern theory of
gravitation and permits to describe successfully different gravitating systems with widely changing
values of physical parameters (energy density, pressure etc). At the same time, GR leads to non
satisfactory consequences in the case of gravitating systems at extreme conditions with extremely
high energy densities, pressures, temteratures, where singular states with divergent values of
certain physical parameters appear. Corresponding problem of GR -- the problem of gravitational
singularities -- was widely discussed in literature [1]. One of the most important cases of this
problem is the problem of cosmological singularity (PCS). All Friedmannian homogeneous isotropic
cosmological models (HICM) of flat and open type and the most part of closed models possess
singular state in the beginning of cosmological expansion, where the value of the scale factor
$R(t)$ of Robertson-Walker metrics vanishes, that limits their existence in the past. According to
wide known opinion, the solution of PCS and generally of the problem of gravitational singularities
of GR has to be connected with quantum gravitational effects, which must be essential at Planckian
conditions, when energy density is comparable with the Planckian one. Previously some regular
bouncing cosmological solutions were obtained in the frame of candidates to quantum gravitation
theory -- string theory/M-theory and loop quantum gravity (see, for example, [2--4]). From physical
point of view, these solutions have some difficulties [5]. In the case of bouncing cosmological
solutions built in the frame of  string theory the condition of energy density positivity for
gravitating matter is violated. In the case of loop quantum cosmology a bounce takes place for
microscopic model having a volume comparable with the Planckian one. If one supposes that the
Universe at compression stage is macroscopic object, one has to explain the transformation of
macro-universe into micro-universe before a bounce. This means one has to introduce some physically
non realistic model, which has to be inverse with respect to inflation.

As it was shown in a number of our papers (see [5-7] and Refs herein), gravitation theory in
4-dimensional physical Riemann-Cartan space-time leads to the solution of PCS and permits to build
satisfactory regular Big Bang theory. Regular character of all solutions  for HICM in this theory
is ensured by gravitational repulsion effect at extreme conditions, when the energy density and the
pressure of gravitating matter are extremely high (although their values can be essentially less,
than the Planckian ones). The principal role at extreme conditions in such theory plays the
space-time torsion. The use of Riemann-Cartan geometry in gravitation theory is motivated by
applying the local gauge invariance principle in theory of  gravitational interaction. The present
paper is devoted to discussion of these problems. In Section 2 some principal ideas of gauge
approach in gravitation theory are presented. In Section 3 the most important results of regular
cosmology built in 4-dimensional Riemann-Cartan space-time are given.

\section{Gauge approach to gravitational interaction and gravitation theory in Riemann-Cartan space-time}

As it is known, the local gauge invariance principle is the basis of modern theory of fundamental
physical interactions. The theory of electro-week interaction, quantum chromodynamics, Grand
Unified models of particle physics were built by using this principle. From physical point of view,
the local gauge invariance principle establishes the correspondence between certain important
conserving physical quantities, connected according to the Noether's theorem with some symmetries
groups, and fundamental physical fields, which have as a source corresponding physical quantities
and play the role of carriers of fundamental physical interactions. The applying of this principle
to gravitational interaction leads, generally speaking, to generalization of Einsteinian theory of
gravitation.

At first time the local gauge invariance principle was applied in order to build the gravitation
theory by Utiyama in Ref.[8] by considering the Lorentz group as gauge group corresponding to
gravitational interaction. Utiyama introduced the Lorentz gauge field, which has transformation
properties of anholonomic Lorentz connection. By identifying this field with anholonomic connection
of riemannian space-time, Utiyama obtained Einsteinian thery of gravitation by this way. The work
by Utiyama [8] was criticized by many authors. At first of all, if anholonomic Lorentz connection
is considered as independent gauge field, it can be identified with a connection of Riemann-Cartan
continuum with torsion, but not riemannian connection  [9-11]. Moreover, if a source of
gravitational field includes the energy-momentum tensor of gravitating matter, we can not consider
the Lorentz group as gauge group corresponding to gravitational interaction. Note that metric
theories of gravitation in 4-dimensional pseudo-riemannian space--time  including GR, in the frame
of which the energy-momentum tensor is a source of gravitational field, can be introduced in the
frame of gauge approach by the localization of 4-parametric translation group [12, 13]
\footnote{Because in the frame of gauge approach the gravitational interaction is connected with
space-time transformations, the gauge treatment to gravitation has essential differences in
comparison with Yang-Mills fields connected with internal symmetries groups. As a result, there are
different gauge treatments to gravitational interaction not detailed in this paper.}. By localizing
4-translations and introducing gauge field as symmetric tensor field of second order, the structure
of initial flat space-time changes, and gauge field becomes to connected with metric tensor of
physical space-time. Because the localized translation group leads us to general coordinate
transformations, from this point of view the general covariance of GR plays the dynamical role. At
the same time the Lorentz group (group of tetrad Lorentz transformations) in GR and other metric
theories of gravitation does not play any dynamical role from the point of view of gauge approach,
because corresponding Noether's invariant in these theories is identically equal to zero [14]. The
other treatment to localization of translation group was presented in [15, 16], where gravitation
field was introduced as tetrad field in 4-dimensional space-time with absolute parallelism. This
theory is not covariant with respect to localized tetrad Lorentz transformations, and in fact it is
intermediate step to gravitation theory with independent gauge Lorentz field. If one means that the
Lorentz group plays the dynamical role in the gauge field theory and the Lorentz gauge field exists
in the nature, in this case we obtain with necessity the gravitation theory in the Riemann-Cartan
space-time (see, for example, [17-19]). Corresponding theory is known as Poincare gauge theory of
gravitation (PGTG). Gravitational field variables in PGTG are the tetrad $h^i{}_\mu$ (translational
gauge field) and the Lorentz connection $A^{ik}{}_\mu$ (Lorentz gauge field); corresponding field
strengths are the torsion tensor $S^i{}_{\mu\nu}$ and the curvature tensor $F^{ik}{}_{\mu\nu}$
defined as
\[
S^i _{\mu \,\nu }  = \partial _{[\nu } \,h^i _{\mu ]}  - h_{k[\mu } A^{ik} _{\nu ]}\,,
\]
\[
F^{ik} _{\mu \,\nu }  = 2\partial _{[\mu } A^{ik} _{\nu ]}  + 2A^{il} _{[\mu } A^k _{|l\,|\nu ]}\,,
\]
where holonomic and anholonomic space-time coordinates are denoted by means of greek and latin
indices respectively. As sources of gravitational field in PGTG are energy-momentum and spin
tensors. The simplest PGTG is the Einstein-Cartan theory based on gravitational Lagrangian in the
form of scalar curvature of Riemann-Cartan space-time [10,11,23]. In certain sense the
Einstein-Cartan theory of gravitation is degenerate gauge theory, in the frame of which the torsion
is connected linearly with spin tensor of gravitating matter and in the case of spinless matter the
torsion vanishes, although the torsion is a gravitational field strength corresponding to localized
translation group. Like gauge Yang-Mills fields, gravitational Lagrangian of PGTG has to include
invariants quadratic in gravitational field strengths - curvature and torsion tensors. The
including of linear in curvature term (scalar curvature) to gravitational Lagrangian is necessary
to satisfy the correspondence principle with GR.

We will consider the PGTG with gravitational Lagrangian given in general form containing different
invariants quadratic in the curvature and torsion tensors
\begin{eqnarray}\label{1}%\fl
{\cal L}_{\rm G}=  f_0\, F+F^{\alpha\beta\mu\nu}\left(f_1\:F_{\alpha\beta\mu\nu}+f_2\:
F_{\alpha\mu\beta\nu}+f_3\:F_{\mu\nu\alpha\beta}\right)+ F^{\mu\nu}\left(f_4\:F_{\mu\nu}+f_5\:
F_{\nu\mu}\right)+f_6\:F^2 \nonumber \\
+S^{\alpha\mu\nu}\left(a_1\:S_{\alpha\mu\nu}+a_2\: S_{\nu\mu\alpha}\right)
+a_3\:S^\alpha{}_{\mu\alpha}S_\beta{}^{\mu\beta}, %\nonumber
\end{eqnarray}
where $F_{\mu\nu}=F^{\alpha}{}_{\mu\alpha\nu}$, $F=F^\mu{}_\mu$,  $f_i$ ($i=1,2,\ldots,6$), $a_k$
($k=1,2,3$) are indefinite parameters, $f_0=(16\pi G)^{-1}$, $G$ is Newton's gravitational
constant.

\section{Regular cosmology in Riemann-Cartan space-time}

According to observational data concerning anisotropy of relic radiation, our Universe was
homogeneous and isotropic beginning from initial stages of cosmological expansion. In connection
with this fact, the investigation of HICM is of greatest interest for relativistic cosmology. In
the frame of PGTG homogeneous isotropic models are described in general case by means of three
functions of time: the scale factor of Robertson-Walker metrics $R(t)$ and two torsion functions
$S(t)$ and $\tilde{S}(t$) determining the following components of torsion tensor (with holonomic
indices) [20]: $S^1{}_{10}=S^2{}_{20}=S^3{}_{30}=S(t)$,
$S_{123}=S_{231}=S_{312}=\tilde{S}(t)\frac{R^3r^2}{\sqrt{1-kr^2}}\sin{\theta}$, where spatial
spherical coordinates are used. The functions $S$ and $\tilde{S}$ have different properties with
respect to transformations of spatial inversions, namely, the function $\tilde{S}(t)$ has
pseudoscalar character. In the case $\tilde{S}(t)=0$ HICM in the frame of PGTG were built and
investigated in a number of papers (see [20, 5-7] and references herein).\footnote{Possible role of
the torsion function $\tilde{S}$ is discussed in the talk [21] of this Conference.}. The curvature
tensor in this case has the following non-vanishing components:
$F^{01}{}_{01}=F^{02}{}_{02}=F^{03}{}_{03}\equiv A$ and
$F^{12}{}_{12}=F^{13}{}_{13}=F^{23}{}_{23}\equiv B$ with
\begin{equation}\label{2}
A=\frac{\left(\dot{R}-2RS\right)^{\cdot}}{R},
 \qquad\qquad
B=\frac{k+\left(\dot{R}-2RS\right)^{2}}{R^2},
\end{equation}
and Bianchi identities in this case are reduced to the only relation
\begin{equation}
\label{3} \dot{B}+2H\left(B-A\right)+4AS=0,
\end{equation}
where $H=\frac{\dot{R}}{R}$  is the Hubble parameter, and a dot denotes differentiation with
respect to time.

The system of gravitational equations of PGTG with gravitational Lagrangian (1) is reduced to three
equations, which by using (3) can be written in the following form [20]
\begin{equation}
\label{syst4}
\begin{array}{l}
6f_0B-12f\left(A^2-B^2\right)-3a\left(H-S\right)S=\rho,\\
2f_0\left(2    A+B\right)+4f\left(A^2-B^2\right)-a\left(\dot{S}+HS-S^2\right)=-p,\\
f\left(\dot{A}+\dot{B}\right)+\left[f_0+\frac{1}{8}a+4f
\left(A+B\right)\right]S=0.\\
\end{array}
\end{equation}
 where $f=f_1+\frac{1}{2}f_2+f_3+f_4+f_5+3f_6$, $a=2a_1+a_2+3a_3$,   $\rho$ is the energy density,
$p$ is the pressure and the average of spin distribution of gravitating matter is supposed to be
equal to zero. The system of equations (4) leads to cosmological equations without high derivatives
if $a=0$ [20] (see below). Then we find from (4) the curvature functions $A$ and $B$ and the
torsion $S$ in the following form
\begin{eqnarray}
\label{oldsol5}
A=-{\frac{1}{12f_0}}\,{\frac{\rho+3p-\alpha\left(\rho-3p\right)^2/2}%
{1+\alpha\left(\rho-3p\right)}}\, ,\nonumber\\
B={\frac{1}{6f_0}}\, {\frac{\rho+\alpha\left(\rho
-3p\right)^2/4}{1+\alpha\left(\rho-3p\right)}}\, ,\\
S(t)= -\frac{1}{4}\frac{d}{dt} \ln\left|1+\alpha(\rho-3p)\right| \, ,\nonumber
\end{eqnarray}
where indefinite parameter $\displaystyle \alpha=\frac{f}{3f_0\,^2}$ has inverse dimension of
energy density. By using expressions (2) of curvature functions for homogeneous isotropic
gravitating models and the solution (5) of gravitational equations of PGTG we obtain the following
generalized cosmological Friedmann equations (GCFE)
\begin{equation}
\label{6}
\displaystyle{\frac{k}{R^2}+\left\{\frac{d}{dt}\ln\left[R\sqrt{\left|1+\alpha\left(\rho-
3p\right)\right|}\right]\right\}^2 }\displaystyle{ =\frac{8\pi G}{3}\;\frac{\rho+
\frac{\alpha}{4}\left(\rho-3p\right)^2}{1+\alpha\left(\rho-3p\right)} \, ,}
\end{equation}
\begin{equation}\fl
\label{7}
\displaystyle{R^{-1}\,\frac{d}{dt}\left[\frac{dR}{dt}+R\frac{d}{dt}\left(\ln\sqrt{\left|1+\alpha\left(\rho
-
3p\right)\right|}\right)\right]} %\nonumber
\displaystyle{=-\frac{4\pi G}{3}\;\frac{\rho+3p- \frac{\alpha}{2}\left(\rho-3p\right)^2}{
1+\alpha\left(\rho-3p\right)}\, .}
\end{equation}
 The conservation law in PGTG has usual form
\begin{equation}
\label{8} \dot{\rho}+3H\left(\rho+p\right)=0.
\end{equation}
By using the GCFE we can investigate HICM, if the content of gravitating matter is known. So, in
the case of HICM filled with non-interacting scalar field $\phi$ minimally coupled with gravitation
and gravitating matter with equation of state in general form $p_m=p_m(\rho_m)$, the energy density
$\rho$ and the pressure $p$ take the form
\begin{equation}
\label{9_} \rho=\frac{1}{2}\dot{\phi}^2+V+\rho_m \quad (\rho>0), \quad
p=\frac{1}{2}\dot{\phi}^2-V+p_m,
\end{equation}
where $V=V(\phi)$ is a scalar field potential. By using (9) and the scalar field equation in
homogeneous isotropic space
\begin{equation}
\label{10} \ddot{\phi}+3H\dph=-\frac{\pat V}{\pat\phi}
\end{equation}
we transform the GCFE (6)-(7) to the following form [7]
\begin{multline}\fl
\label{e11}
 \left\{
 H
\left[
 Z+3\alpha
 \left(
   \dph^2+\frac{1}{2} Y
 \right)
\right]
 +3\alpha\frac{\pat V}{\pat\phi}\dph\right\}^2
 +\frac{k}{R^2}\,Z^2 \\ %\nonumber\\
 =\frac{8\pi G}{3}\,
 \left[
   \rhm+\frac{1}{2}\dph^2+V+\frac{1}{4}\alpha\,
   \left(4V-\dph^2+\rhm-3\prm\right)^2
 \right]
\,Z,
\end{multline}
\begin{multline}\fl
\label{e12}%modified
\dot{H}\left[
 Z+3\alpha
 \left(
   \dph^2+\frac{1}{2} Y
 \right)
\right] +3H^2
 \left[
   Z-\alpha\dph^2
   +\alpha Y
   \vphantom{\frac{d^2\prm}{d\rhm^2}}
   \right. \\ %\nonumber\\
   \left.
   -\frac{3\alpha}{2}
        \left(\frac{d\prm}{d\rhm}Y
         +3\left(\rhm+\prm\right)^2 \frac{d^2\prm}{d\rhm^2}
         \right)
 \right]
+3\alpha \left[
    \frac{\pat^2 V}{\pat\phi^2}\dph^2-\left(\frac{\pat V}{\pat\phi}\right)^2
\right] \\ %\nonumber\\
=8\pi G \left[
    V+\frac{1}{2}\left(\rhm-\prm\right)
    +\frac{1}{4}\alpha\left(4V-\dph^2+\rhm-3\prm\right)^2
\right]-\frac{2k}{R^2}Z.
\end{multline}
By using (10)-(12) inflationary cosmological models were investigated in [6].

The principal difference of (6)--(7) from Friedmannian cosmological equations of GR is connected
with terms containing the parameter $\alpha$. These terms arise from quadratic in the curvature
tensor part of gravitational Lagrangian, which unlike metric theories of gravitation does not lead
to high derivatives in cosmological equations. The value of $|\alpha|^{-1}$ determines the scale of
extremely high energy densities. Solutions of GCFE (6)-(7) coincide practically with corresponding
solutions of GR, if the energy density is small $\left|\alpha(\rho-3p)\right|\ll 1$
($p\neq\frac{1}{3}\rho$). The difference between GR and PGTG can be essential at extremely high
energy densities $\left|\alpha(\rho- 3p)\right|\gtrsim 1$. Ultrarelativistic matter
($p=\frac{1}{3}\rho$) and gravitating vacuum ($p=- \rho$) with constant energy density are two
exceptional systems, because  GCFE (6)--(7) are identical to Friedmannian cosmological equations of
GR in these cases independently on values of energy density, and the torsion function vanishes.
Properties of solutions of  equations (6)--(7) at extreme conditions depend on the sign of
parameter $\alpha$ and certain restriction on equation of state of gravitating matter. The study of
inflationary models including scalar fields shows that GCFE (6)--(7) lead to acceptable restriction
for scalar field variables if $\alpha>0$ [5]. In the case $\alpha>0$ all cosmological solutions
have regular bouncing character, if at extreme conditions $p>\frac{1}{3}\rho$. There are physical
reasons to assume, that the restriction $p>\frac{1}{3}\rho$ is valid for gravitating matter at
extreme conditions [22]. Note, that this condition is valid for so-called stiff equation of state
$p=\rho$ used in the theory of the early Universe (Ya.~B.~Zeldovich and others).

The GCFE lead to restrictions on admissible values of energy density. In fact, if the energy
density $\rho$ is positive and $\alpha>0$, from equation (6) in the case $k=+1$, $0$ follows the
relation:
\begin{equation}
\label{13} Z\equiv 1+\alpha\left(\rho-3p\right)\ge 0.
\end{equation}
The condition (13) is valid not only for closed and flat models, but also for cosmological models
of open type ($k=-1$) [5]. In the case of models filled with usual gravitating matter without
scalar fields the equation $Z=0$ determines limiting (maximum) energy density, and regular
transition from compression to expansion (bounce) takes place for all cosmological solutions by
reaching limiting energy density. In the case of systems including also scalar fields a bounce
takes place in points of so-called "bounce surfaces" in space of variables ($\phi$, $\dot\phi$,
$\rhm$) [5]. Near bounce surfaces as well as bounds $Z=0$ gravitational interaction has the
character of repulsion, but not attraction. The domain of energy densities and other variables of
gravitating matter, when gravitational repulsion effect takes place, depends on matter properties
at extreme conditions (equation of state of gravitating matter, scalar field potentials) [7].

As it was noted above, gravitational repultion effect ensures regular behaviour of cosmological
solutions in metrics, Hubble parameter, its time derivative in the frame of classical
field-theoretical description without quantum gravitational corrections. Any cosmological solution
has bouncing character and includes the regular transition from compression to cosmological
expansion. As illustration, below particular cosmological solution for flat inflationary
cosmological model is given (Fig.1 - Fig.3) [5]. The model is filled with scalar field with
potencial $V=\frac{1}{2}m^2\phi^2$ (${m=10^{-6}M_p}$, $M_p$  is the Planckian mass)    and
ultrarelativistic matter. The solution was obtained by numerical integration of Eqs. (10), (12) and
by choosing some initial conditions at a bounce; the value of parameter $\alpha=10^{14}M_p^{-4}$,
Planckian system of units is used in Fig.1 - Fig.3. Note that  duration of the stage of transition
from compression to expansion is extremely small (several order smaller than duration of
inflationary stage). This means, that such models correspond to regular Big Bang or Big Bounce. As
numerical analysis of inflationary cosmological models in PGTG shows [6], quantitative differences
of such models in comparison with similar models in GR are possible at the end of inflationary
stage, if the value of $|\alpha|^{-1}$ is much less than the Planckian energy density. In this case
corresponding corrections concerning anisotropy of relic radiation in considered theory are
possible.
\begin{figure*}[htb!]
\begin{minipage}{0.48\textwidth}%\centering{
\includegraphics{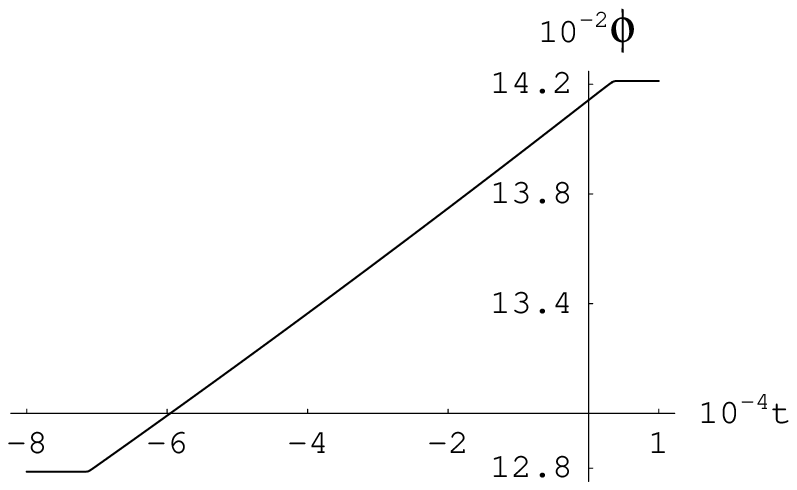}%
\end{minipage}\, \hfill\,
\begin{minipage}{0.48\textwidth}%\centering{
\includegraphics{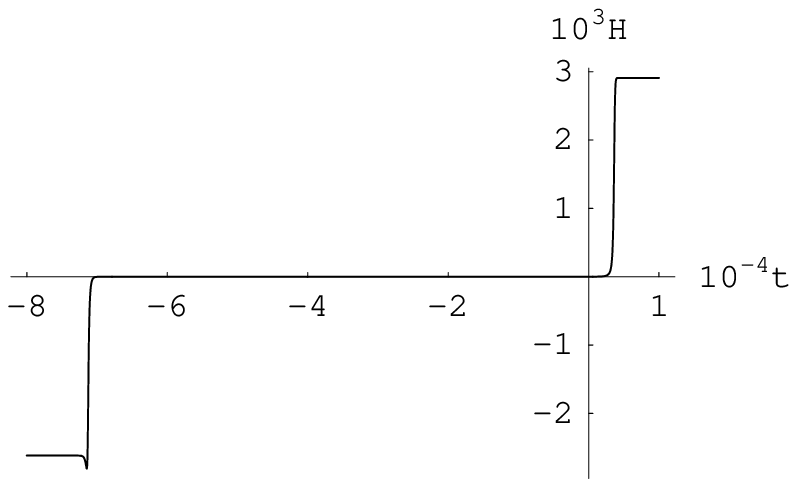}%
\end{minipage}%\\
\caption{\label{figm2}The stage of transition from compression to expansion.}
\end{figure*}

\begin{figure*}[htb!]
\begin{minipage}{0.48\textwidth}%\centering{
\includegraphics{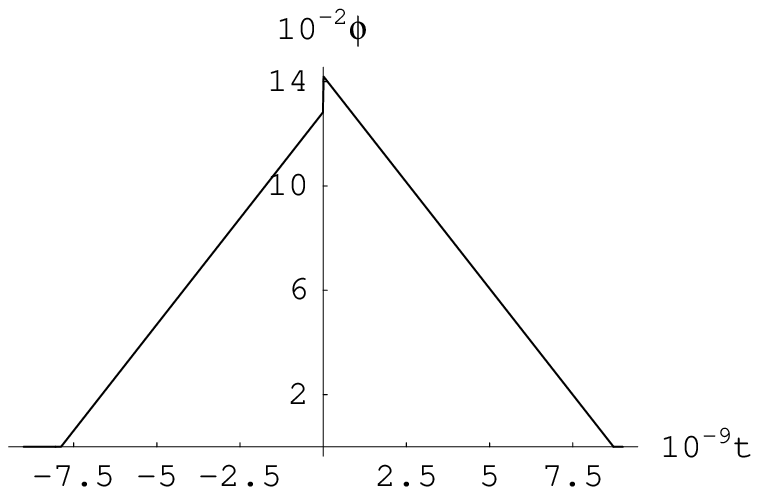}%
\end{minipage}\, \hfill\,
\begin{minipage}{0.48\textwidth}%\centering{
\includegraphics{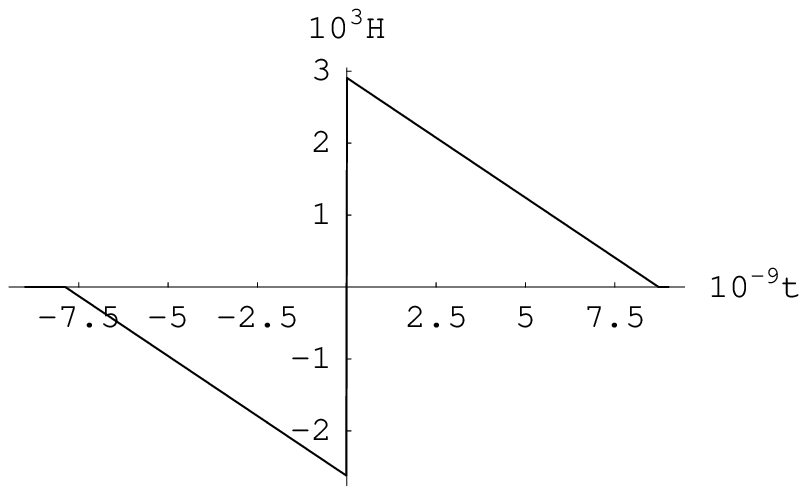}%
\end{minipage}
\caption{\label{figm3}Quasi-de-Sitter stage of compression and inflationary stage.}
\end{figure*}

\begin{figure*}[htb!]
\begin{minipage}{0.48\textwidth}%\centering{
\includegraphics{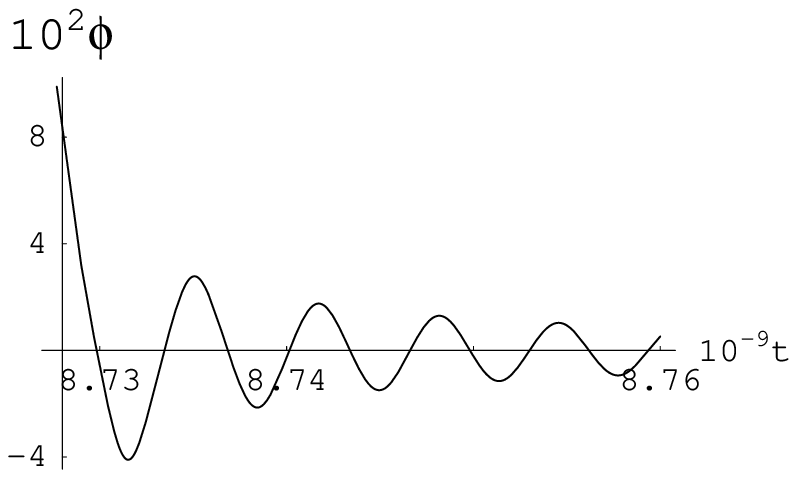}%
\end{minipage}\, \hfill\,
\begin{minipage}{0.48\textwidth}%\centering{
\includegraphics{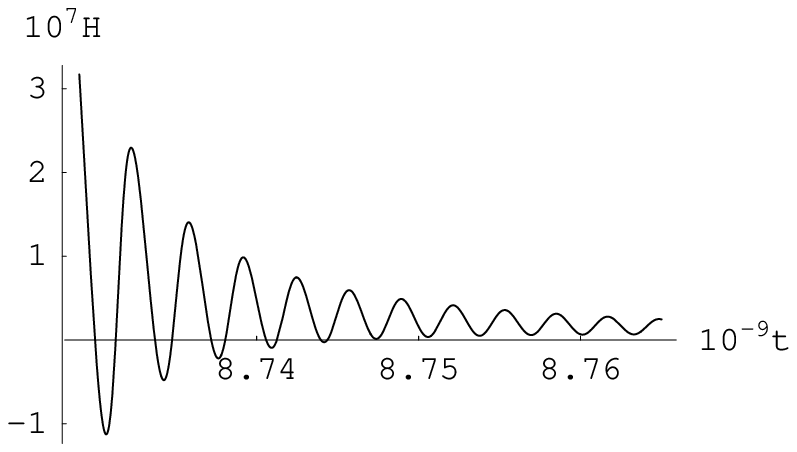}%
\end{minipage}
\caption{\label{figm4}The stage after inflation.}
\end{figure*}

\section{Conclusion}

The analysis of HICM carried out in the frame of PGTG shows, that this theory leads to the solution
of principal problem of GR -- problem of cosmological singularity -- and permits to build regular
cosmology. It is because the gravitational interaction at extreme conditions in PGTG has the
character of repulsion but not attraction. This effect is connected with geometrical structure of
physical space-time in PGTG, namely with the space-time torsion.


\begin{thebibliography}{99}
\bibitem{ml1} Hawking S W and Ellis G F R 1973 {\it The Large Scale Structure of Space-Time\/}
(Cambridge: Cambridge University Press)
\bibitem{ml4} Gasperini M, Veneziano G 2003 {\it Phys.
Rep.\/} {\bf 373} 1 ({\it Preprint\/} hep-th/0207130)
\bibitem{ml5} Bozza V, Veneziano G 2005 Scalar perturbations in regular two-component bouncing
cosmologies {\it Preprint\/} hep-th/0502047; 2005 Regular two-component bouncing cosmologies and
perturbations therein {\it Preprint\/} gr-qc/0506040
\bibitem{ml6} Bojowald M 2002 {\it Class. Quant. Grav.\/} {\bf 19} 2717 ({\it Preprint\/} gr-qc/0202077)
\bibitem{ml7} Minkevich A V 2006 {\it Gravitation\&Cosmology} {\bf 12}  no.~1(45) 11  ({\it
Preprint\/} gr-qc/0506140)
\bibitem{ml17} Minkevich A V and Garkun A S 2006 {\it Class. Quantum Grav.\/} {\bf 23} 4237 ({\it
Preprint\/} gr-qc/0512130)
\bibitem{mla7} Minkevich A V On gravitational repulsion effect at
extreme conditions in gauge theories of gravity, {\it Preprint\/} gr-qc/0512123 (to be published in
Acta Phys. Polon. {\bf B}, 2007).
\bibitem{mla8} Utiyama R 1956 {\it Phys. Rev.\/} {\bf 101} 1597
\bibitem{mla9} Brodskii A M, Ivanenko
D, Sokolik H A 1961 {\it Zhurnal Eksp. Teor. Fiz.\/} {\bf 41} 1307; 1962 {\it Acta Phys. Hungar.\/}
{\bf 14} 21
\bibitem{mla10} Kibble T W B 1961 {\it J. Math. Phys.\/} {\bf 2} 212
\bibitem{mla11} Sciama D W 1962 In: {\it Recent Developments in GR\/} (Warsaw-New York:
Pergamon Press and PMN) 321
\bibitem{mc16} Minkevich A V 1966 {\it Vestsi Akad. Nauk BSSR.\/} Ser. fiz.-mat., no.~4, 117
\bibitem{mla13} Utiyama R, Fukuyama T 1971 {\it Progr. Theor. Phys.\/} {\bf 45} 612
\bibitem{mc17} Minkevich A V, Kudin V I 1974 {\it Acta Phys. Polon.\/} {\bf B5} 335
\bibitem{mla15} Hayashi K, Nakano T 1967 {\it Progr. Theor. Phys.\/} {\bf 38} 491
\bibitem{mla16} Hayashi K, Shirafudji T 1979 {\it Phys. Rev.\/} {\bf D19} 3524
\bibitem{mla17} Cho J M 1976 {\it Phys. Rev.\/} {\bf D14} 3335
\bibitem{hehl} Hehl F W 1980 {\it in:\/} ``Cosmology and Gravitation'' (New York: Plenum Press)
\bibitem{hay} Hayashi K, Shirafuji T 1980 {\it Progr. Theor. Phys.\/} {\bf 64} 866;
{\bf 64} 1435; {\bf 64} 2222
\bibitem{ml10} Minkevich A V 1980 {\it Vestsi Akad. Nauk BSSR. Ser. fiz.-mat.\/} no.~2 87;
{\it Phys. Lett.\/} A {\bf 80} 232
\bibitem{mla21} Minkevich A V, Garkun A S and Kudin V I 2006
Homogeneous isotropic cosmological models with pseudoscalar torsion function in Poincare gauge
theory of gravity and accelerating Universe. Proc. of the 5th Intern. Conf.
Boyai-Gauss-Lobachevsky: Non-Euclidean Geometry in Modern Physics, Minsk, Belarus, Oct. 10-13,
2006, pp. 196--202.
\bibitem{ml14} Zeldovich Ya B, Novikov I D  1975 {\it Structure and evolution of the Universe\/}
(Moscow: Nauka) 37
\bibitem{traut} Trautman A 1973 {\it Nature (Phys. Sci.)\/} {\bf 242}, 7 \end{thebibliography}
\end{document}